# Reflections on Disentanglement and the Latent Space

Ludovica Schaerf[1,2]

[1] University of Zurich, Zurich, Switzerland
[2] Max Planck Society, Rome, Italy
ludovica.schaerf@uzh.ch

**Abstract**

The latent space of image generative models is a multi-dimensional space of compressed hidden visual knowledge. Its entity captivates computer scientists, digital artists, and media scholars alike. Latent space has become an aesthetic category in AI art, inspiring artistic techniques such as the latent space walk, exemplified by the works of Mario Klingemann and others. It is also viewed as cultural snapshots, encoding rich representations of our visual world. This paper proposes a double view of the latent space, as a multi-dimensional archive of culture and as a multi-dimensional space of potentiality. The paper discusses disentanglement as a method to elucidate the double nature of the space and as an interpretative direction to exploit its organization in human terms. The paper compares the role of disentanglement as potentiality to that of conditioning, as imagination, and confronts this interpretation with the philosophy of Deleuzian potentiality and Hume's imagination. Lastly, this paper notes the difference between traditional generative models and recent architectures.

## 1 Introduction

The year 2023 bore witness to rapid and pervading improvements in generative models, including image generation models pioneered by Dall-E, Midjourney, and Stable Diffusion (Rombach et al. 2022; Ramesh et al. 2021). These models learn from visual inputs to reconstruct and recreate images and condition their features using external modalities, steering the final image. The popularity of these methods notably increased with the introduction of *language conditioning* (Radford et al. 2021).

A widespread strand of image generative models (IGM) features a *latent space* in its architecture (i.e. autoencoders, generative adversarial networks, diffusion). A latent space is an abstract multi-dimensional space containing compressed feature values that we cannot, in most cases, interpret directly[1]. Latent space is juxtaposed to an observable space in statistical terminology. It is not directly inspectable, but it describes the original data with a certain precision. Within the architecture, it is the space in which the data lies in the *bottleneck layer*[2]—the layer with the fewest neurons (in Fig. 1).

This space forms a stimulating entity for computer scientists, digital artists, and media scholars alike. The popular artist studio, Refik Anadol Studio (RAS), imagines latent space as a process of dreaming and hallucinating of digested and condensed memories[3]. In this metaphor, the memories are the data, whereas the dreams are the generations of the model.

---

[1] Definition adapted from this thread
[2] Formulation found in this blog

Latency is an aesthetic category for many multimedia artists. The *latent space walk* is a topical technique in AI art, where artists take a model and a starting point within its latent space and move in random or controlled directions from that point to generate a pseudo video. Instances of this exploration are Memories of Passersby I by Mario Klingemann (Klingemann, 2018), Anna Ridler's Mosaic Virus[4], Sofia Crespo's Neural Zoo[5], and many others. These videos explore the probabilistic distribution within the latent space, which yields a continuous and semantically[6] meaningful space.

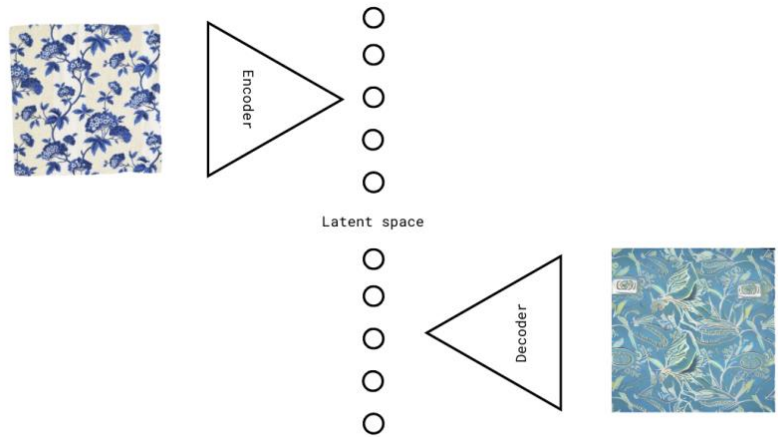

**Fig. 1.** Schematization of an IGM architecture with latent space.

From the powerful generative capacity of these models, we can hypothesize that they have learned to represent and reuse considerable visual knowledge. This knowledge may differ from human processing. For this reason, researchers have started to identify and analyze these models as *cultural snapshots* or cultural objects (Cetinic 2022; Rodríguez-Ortega 2022; Impett and Offert 2023) that contain within their weights and connections their cultural representation of the data. These representations are hidden in the latent space.

The extraction of an interpretable representation within high-dimensional data, in the computer vision domain, is called disentanglement. *Weakly-supervised disentanglement*, accordingly, is a method to find multi-dimensional representations of features known in the data, such as the main color of an image, within the latent spaces of a model (Locatello et al. 2020).

The remainder of this essay explores two conceptual interpretations of the latent space, as a *multi-dimensional archive of culture* and as a *multi-dimensional space of potentiality*, and proposes disentanglement as a technical methodology to shed light on the space.

Broadly speaking, it addresses the following questions:

---

[3] https://www.goethe.de/prj/k40/en/kun/lat.html
[4] https://annaridler.com/mosaic-virus
[5] https://neuralzoo.com/
[6] It is to be noted that *semantic* in computer vision is used as a vague term referring to the content of the images.

- How can theories from the humanities help us understand the concepts and affordances of (disentangled) latent spaces?
- How can technical methods, such as disentanglement, help us explore and support humanist theories of latent spaces?

This essay proposes several observations linked by the common theme of (disentangled) latent space and offers an ensemble of preliminary thoughts with the intention of stimulating further reflections in the reader.

## 2 Latent Space Interpretations

Much like light is both a particle and a wave, a latent space can be technically viewed in two ways.

The first, hereafter referred to as the *code view*, is the interpretation of a latent space as the values that each neuron takes within the bottleneck layer. These values form the code that, when passed through the decoder, will produce an image. The ensemble of the values of the neurons creates a code that determines, one-to-one, a resulting image – allowing for some small randomness in certain models. We analyze this view in connection to the concept of the archive and interpret the latent space as a multi-dimensional archive of culture, where each image, existing or potential, is given a code that allows an organization in terms of visual similarity.

The ensemble of neurons in the bottleneck creates a multi-dimensional space, whose dimensionality corresponds to the number of neurons. A code is therefore a point within this space. Analyzing the latent space in these terms is what we will call the *space view*. The space generated by these models is often continuous, which means that between any two images, there is an infinite number of images. Unlike traditional archives, a continuous space allows representing not only what exists, but all the potential states in between. Therefore, we interpret the latent space in this view as a multi-dimensional space of potentiality.

In the following subsections, we present an attempt at defining the two interpretations and expand on their conceptualization.

### 2.1 Multi-Dimensional Archive of Culture

The latent space is:

> a multi-dimensional and constructed repository of *potential and actual* cultural artifacts, memories, and knowledge, governed by specific rules of *visual similarity* [in the IGM context]. It serves as a site for storing and generating, while also *exhibiting the cultural biases* inherent to the choice of data and the technical architecture of the machine.

In The Archaeology of Knowledge, Foucault defines an archive as 'the first the law of what can be said, the system that governs the appearance of statements as unique events', where the objects are ordered meaningfully according to multiple relationships (Foucault 2012). The codes of the latent space order the unique artifacts in a system of complex relationships using constructed, implicit, rules of similarity. Every object that can be ingested in the model—if there exists a code that can generate the object (in technical terms if its mode does not collapse)—is included in the repository of the model. The 'mode collapse'[7], typical of GAN models, determines the exclusion of some visual elements, determining a system of memory and forgetting akin to Derrida's Archive Fever (Derrida 1996).

---

[7] For a definition of mode collapse refer [here](#).

Similarly to other archives, the cultural identity of a model is created by the data they are fed and is re-elaborated and assimilated as a representation within the model (Fazi 2021). The representation created by the model, if this is abstract enough (Bengio 2013)[8], contains cultural data that is unfolded into knowledge. Therefore, the model can be seen as a cultural snapshot of its data re-elaborated into machinic knowledge.

Pasquinelli and Joler, 2021, offer the metaphor of machinic knowledge as knowledge passed through diffracted lenses, where information is remixed with a certain distortion by the architecture and the training procedure of the model (Pasquinelli and Joler 2021). While distortions and selection are innate in any archive, in the latent space archive, one can "'search' for images that don't exist yet and therefore have never been indexed" before (Meyer 2023).

### 3.2 Multi-Dimensional Space of Potentiality

The latent space is also:

> a multidimensional space of potentiality, a *mathematic generalization* representing the ensemble of potential realities learned through the data. The space statistically fits the cultural artifacts into a *continuous* and fluid space. New cultural artifacts can emerge through the *creative synthesis* and transformation of these potentials.

According to Rodríguez-Ortega, nowadays cultural objects are embedded and transformed into n-dimensional spaces, which are created and can be analyzed through computation (Rodríguez-Ortega 2022). These spaces provide a conceptual framework for thinking of cultural objects.

While each model is inherently different, they all learn a latent representation to best fit the training data distribution. Because the spaces we are dealing with in this paper are continuous spaces that statistically represent the data, the holes left by the discrete points of the training data are filled with what is statistically possible. Statistical possibility of the latent space is a concept that is well explained by the philosophy of potentiality. In Deleuzian terms, *potentiality* is an existence that has not yet become actual, in his terms, *virtuality*. Differently from *possibility*, which can be realized, a point in the latent space already exists, it only needs to be actualized - decoded into an image (Deleuze 2014; "An Analysis of Deleuze's Concept of Potentiality" 2023). The decoding instantiates change, transforming potentiality into *actuality*, as in an Aristotelic Hylomorphism (Simpson 2023). The space itself determines a plane of immanence, representing the ensemble of potential – or virtual – states of reality.

---

[8] Human reasoning follows certain abstraction created in our minds, as such, we are more suited to understand representations within models if those form the same level of abstractions. (Bengio 2013)

# 3 Disentanglement

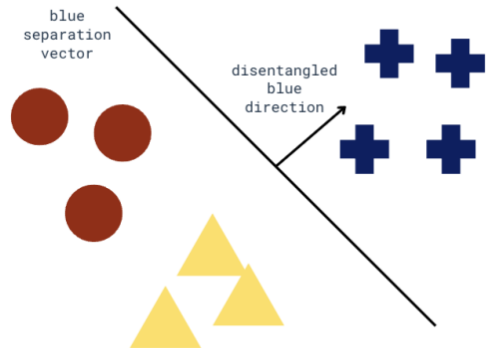

**Fig. 2.** Schematization of semi-supervised disentanglement (InterfaceGAN).

The affordances of this ambivalent space are significant and explain its popularity in fields outside of computer science. Nonetheless, its organization into an archive and a space follows rules that evade human cognition. As a prerogative of any deep learning model, the features that determine the organization of the latent space are implicitly learned by the model during training. This begs the question, how do we take advantage of this complex entity?

A research line that investigates the potential interpretation of latent spaces is that of *disentanglement*. Disentanglement is defined in computer vision as the task of learning or identifying traversal directions in the latent space of models controlling only one factor of variation (Choi et al. 2022). This implies that – in certain models – we can find a representation of a factor of variation that manipulates only a desired feature. This direction lies in the n-dimensional space and allows visual modifications to the output image toward the defined feature. As such, disentanglement allows exploring the latent space through an interpretative direction.

There exist two general types of disentanglement: unsupervised and semi-supervised. The first includes methods such as GANSpace and extracts representations that are most important within the space but are not directly humanly interpretable (Härkönen et al. 2020). The second finds representations based on features that are known in the data. Popular methods for weakly-supervised disentanglement include InterfaceGAN, schematized in Fig. 2 (Shen et al. 2022). Given a feature of interest, in our example, the main color of the resulting image being blue, we can determine which points of the latent space yield which main color. With this information, we can find a separation vector in the latent space using boundary-based machine learning methods (support vector machines, logistic regression). The direction exiting that separation can be interpreted as the representation of the feature. In practice, this means that if we add that direction vector to any point in the latent space we will get an image that is bluer than the previous.

## 3.1 Disentangled Latent Spaces

Despite originating as a purely technical field, semi-supervised disentanglement serves an important role in the latent space.

In the code view, it allows "searching" for instances using human criteria of organization. We can disentangle the country where an image is taken, and sample or modify codes along this disentangled direction yielding images that follow this organization. Disentanglement can be done with any metadata that exists or can be extracted from an image. Differently from a normal archive, the disentangled direction will only work to the extent that the model has implicitly extracted that feature. That is, if we can successfully disentangle a feature, it means that the model has learnt its distinctiveness. Therefore, only information that is transformed into machinic knowledge can be disentangled.

In the space view, disentanglement allows exploring the *potential* states of the model through an interpretative lens. In the previous example, it would answer the question: what can be from that country? Or how would this image be if it were from that country? The model would be merely navigating the cultural snapshot it created. In practical terms, we would expect the model to change only to the extent that it *makes sense* within its representation of the data.

### 3.4 The Role of Conditioning

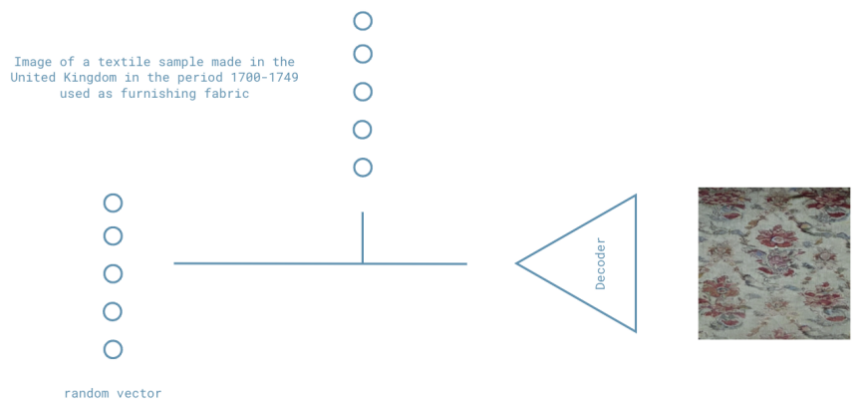

**Fig. 3.** Schematization of conditioning.

As an image editing tool, disentanglement is often compared to conditioning. Conditioning in computer vision refers to the process of incorporating additional information into a model to influence its output (Mirza and Osindero 2014). For example, in image generation tasks, conditions can include textual descriptions, class labels, or other auxiliary data that guide the model to produce images that match the specified criteria. Conditioning evolved from explicit classifiers to concatenation to the input or latent space, to insertion through attention mechanisms to classifier-free guidance. While conditioning includes a breadth of methods, the common feature between all these techniques is the insertion of external knowledge, as seen in Fig. 3. If on one side, disentanglement exploits the potential states of the model itself, conditioning enlarges the possibilities of the model by training alongside an external input. Text-to-image (TTI) models are a popular instance of this. We can prompt a model with 'An astronaut riding a horse' and the model would generate the resulting image despite it not existing in the *virtual*.

This process goes in the direction of how the empiricist philosopher Hume discusses the *imaginative* faculty of humans. When compounding an astronaut and a horse, we divide, compose, or augment simple ideas to synthesize complex, novel ideas (Hume et al., 2007). In short, looking at original images is looking at images in the space of *actuality*, and looking at disentangled codes is looking at images in the space of *potentiality*, and looking at generated conditioned images is looking at images in the space of *imagination*[9].

I believe this imaginative ability of compounding ideas emerges from the compounding of modalities that take place in conditioning. In this case, an astronaut on a horse is possible in the textual modality, and its possibility is translated to the visual realm when training on conditioning.

### 3.5 GANs and/or Diffusion?

Most of the popularity of latent space and disentanglement originated with the invention of generative adversarial networks (GANs). These models comprise a generator and a discriminator network. The generator takes a random multi-dimensional vector (the latent space) and decodes it into an image. The discriminator tries to determine whether the image generated is real. The goal of the adversarial process is for the generator to fool the discriminator. In this case, the original latent space has a comparatively low dimensional latent space, with the most common dimensionality, around 512 neurons (Karras, Laine, and Aila 2019). This space is known to have impressive semantic qualities, including notable disentanglement capabilities. This space is univocal in that for a chosen bottleneck layer that one latent space determines the generation of an image. It is one-dimensional, as it has 512 neurons for depth and only one for width. It is non-spatial because of the unit width. It does not create a situation where certain neurons only attend to some parts of an image. Because of all these features, the space is sufficiently abstract to encapsulate all the properties mentioned thus far.

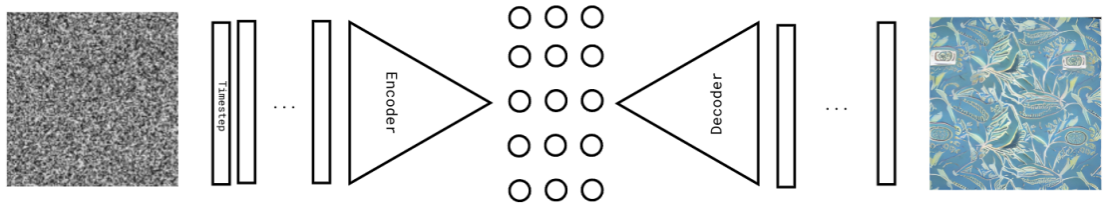

**Fig. 3.** Schematization of diffusion models.

The popularity of diffusion models begs the question of whether they feature a semantic disentangled latent space. Technical research has shown that the h-space (the dots in Fig. 4) demonstrates sufficient evidence to be considered a semantic space that can be disentangled (Kwon, Jeong, and Uh 2023). The concept of 'semantic' is however ill-defined. Schaerf et al., dealing with representations of colors within GAN and diffusion models, show that while the similarity of the representations of the colors in the latent space of StyleGAN mirrors the similarities in human perception and color encoding, the latent spaces of diffusion do not present the same property, suggesting a non-semantic organization of the space (Schaerf, Alfarano, and Postma 2024). This may be caused by a number of reasons. Notably, diffusion models are trained to iteratively model away the noise from an image. Because of the iterative procedure, the latent space is not univocal, rather there is a latent space for every timestep. Furthermore, the dimensionality of the aforementioned h-space is, usually, 1280x8x8 (81920) (Rombach et al. 2022). This is more than 100 times the space of GANs, and is therefore likely rather sparse. Finally, it still presents spatial information, it can be interpreted as 1280 vectors of 8x8 images.

---

[9] Correspondences to latent space imagination and possibility can be found in Margaret Boden's types of creativity, as detailed in Schaerf et al. (2024).

While this research is very preliminary, we see an evolution of IGM away from traditional latent spaces or completely missing a latent space. Therefore, new theories will need to be developed for the rising architectures.

## 4 Conclusion

This paper discussed several ideas on the latent space and the role of disentanglement. Drawing from the importance of the latent space in arts and media studies, we attempted to delve into views, frameworks, and methods of the latent space.

The paper presented the latent space as a double object, an archive of culture, and a space of potentiality. It confronted our interpretations with the philosophical concepts of potentiality and imagination, and later introduced disentanglement as an avenue to represent features to make sense of the organization of the archive and space. Lastly, the text discussed the difference between disentangled and conditioning, proposed as a space of imagination, and proposed some differences between traditional latent space and current innovations.

This paper constitutes just the beginning of a reflection, and its aim is merely to stimulate future considerations.

## Acknowledgments

This paper could not have been written without the useful comments from Dario Negueruela del Castillo and Valentine Bernasconi, all the mentors and peers of the School of X, particularly Angela Ferraiolo, Caterina Moruzzi, Grace Han, and Veronica Silva, and the interesting conversations on the latent space as modern archive with Nuria Rodríguez-Ortega.